\documentclass[12pt]{article}
\usepackage{graphicx}
\usepackage{bm}
\usepackage{rotating}
\usepackage{color}
\begin{document}
\textheight=23.0cm
\begin{center}
{\large Pressure induced phase transitions in PbTiO$_3$ - a query for the polarization rotation theory}\\
J. Frantti, Y. Fujioka and R. M. Nieminen\\
Laboratory of Physics, Helsinki University of Technology, P.O. Box 4100, FIN-02015 HUT, Finland\\
\end{center}
%\footnote[1]{To whom correspondence should be addressed. E-mail: jfr@fyslab.hut.fi}
\begin{abstract}
%Lead titanate (PbTiO$_3$) is a classical ferroelectric perovskite possessing exceptionally 
%large spontaneous electric polarization and electromechanical coupling coefficients. Structural 
%changes occurring in PbTiO$_3$ as a function of temperature are well known: there is only one 
%phase transition at 493$^{\circ}$C at which symmetry is lowered from the cubic $Pm\bar{3}m$ to 
%tetragonal $P4mm$. To understand the piezoelectric properties of PbTiO$_3$ its behaviour under 
%high pressure is clearly of central importance. Despite the nominally simple crystal structure 
%the phase transition sequence and functional properties under large stress are still not well 
%understood. Ferroelectric properties strongly depend on the primitive cell volume. 
Our first-principles computations show that the ground state of PbTiO$_3$ under hydrostatic pressure 
transforms discontinuously from $P4mm$ to $R3c$ at 9 GPa. Spontaneous polarization decreases with 
increasing pressure so that the $R3c$ phase transforms to the centrosymmetric $R\bar{3}c$ phase 
at around 30 GPa. The first-order phase transition between tetragonal and rhombohedral phase is 
exceptional since there is no evidence for a bridging phase. The essential feature of the $R3c$ and 
$R\bar{3}c$ phases is that they allow the oxygen octahedron to increase its volume $V_B$ at the 
expense of cuboctahedral volume $V_A$ around a Pb ion. This is further supported by the fact that 
neither the $R3m$ nor $Cm$ phase, which keep the $V_A/V_B$ ratio constant, is a ground state 
within the pressure range between 0 and 40 GPa. Thus tetragonal strain is dominant up to 9 GPa, 
whereas at higher pressures efficient compression through oxygen octahedra tilting plays the 
central role for PbTiO$_3$. Previously predicted pressure induced colossal 
enhancement of piezoelectricity in PbTiO$_3$ corresponds to unstable $Cm$ and $R3m$ phases. This 
suggests that the phase instability, in contrast to the polarization rotation, is responsible for 
the large piezoelectric properties observed in systems like Pb(Zr,Ti)O$_3$ in the 
vicinity of the morphotropic phase boundary.

\end{abstract}

%\pacs{77.84.Dy  61.12.Ld  61.50.Ah  81.30.Dz}
%\end{frontmatter}

%\maketitle

%\section{Introduction}
In ferroelectric crystals electric polarization occurs spontaneously. The spontaneous 
polarization occurs when the energy gained by the interaction of the dipoles induced by the local 
field is larger than the energy required to produce the dipoles \cite{Levy}. This viewpoint was adopted  
once early attempts for explaining ferroelectricity were developed: the ferroelectric state is stable when the 
attractive dipolar energy exceeds the distortion energy. It soon turned out that the model in its simplest form
was insufficient. For BaTiO$_3$ a phenomenological model, based on Slater's 'rattling' Ti ion and the fact 
that the environment of the oxygen ions is noncubic, was developed to explain ferroelectricity. A rattling Ti ion 
creates large ionic polarizability (implying small distortion energy), whereas the noncubic environment 
of oxygen ions increases the local field contribution from the titanium. A large local field implies large 
dipolar attraction energy dominating over the distortion energy. Quantitative treatment of ferroelectricity 
in BaTiO$_3$ and PbTiO$_3$ was provided by first principles computations which, instead of a rattling titanium ion, 
considered the hybridization between Ti $3d$ and oxygen $2p$ states \cite{Cohen}. These computations were able to 
explain the difference between the zero pressure ground states of BaTiO$_3$ (rhombohedral) and PbTiO$_3$ 
(tetragonal) by considering Ba essentially as a closed shell ion and taking the hybridization of Pb $6s$ 
and oxygen $2p$ states into account. The ground state is very sensitive against primitive cell volume and one may 
expect significant changes in the structure of PbTiO$_3$ as a function of hydrostatic pressure. Recent development 
on ferroelectrics are due to the density functional theory (DFT) which provides an accurate value 
for the total energy except for the errors due to the approximate forms used to describe the exchange-correlation 
functional. 
%Local density approximation (LDA) is commonly applied in the case of ferroelectrics though it tends to 
%overbind the atoms and thus underestimates the primitive cell volume and phase transition 
%pressures. 

We used the ABINIT DFT code to compute the total energies at different pressures \cite{Abinit}. The computations 
were carried out within the local-density approximation (LDA) approximation and a plane wave basis. Norm-conserving 
pseudo-potentials (scalar-relativistic for Pb and non-relativistic for Ti and O \cite{Grinberg}) were generated using the 
OPIUM pseudopotential generator package \cite{Opium}. The 
valence states included $5d6s6p$ (Pb), $3s3p3d4s4p$ (Ti) and $2s2p3d$ (O) states. The energy cutoff was 75 hartree and the 
$k$-point meshes were $6\times 6\times 6$, except for the $R3c$ and $R\bar{3}c$ symmetries, for which it was 
$4\times 4\times 4$ (due to the larger primitive cell), and $I4cm$ symmetry, for which a mesh $6\times 6\times 4$ was 
used. Denser $k$-point meshes were used for computing spontaneous polarization through Berry's phase method. 
Structural optimization was carried out till force acting on each atom was below $5\times 10^{-6}$ eV/\AA.
Phonon frequencies and the corresponding atomic displacements (eigenvectors) were computed through response function 
method implemented in ABINIT. 
%Real prediction of a ground state at each pressure is a very challenging task. 
To tackle the problem of structural optimization we selected various potential crystal symmetries for which the enthalpies 
were computed as a function of pressure. Thus, due to the polarization rotation theory \cite{Fu} the case where the $Cm$ 
symmetry bridges the tetragonal $P4mm$ and rhombohedral $R3m$ symmetries is a particularly attractive possibility to be examined. 
The $Cm$ phase is a common subgroup of the $P4mm$ and $R3m$ according to the group-subgroup chains 
$P4mm \rightarrow Cmm2 \rightarrow Cm$ and $R3m \rightarrow Cm$. In such a case a large electromechanical coupling coefficient, 
assigned to the $Cm$ phase, was predicted \cite{Wu}. This was based on the idea that the $Cm$ phase would allow energetically 
favourable continuous rotation of polarization between the $[ 001 ]$ ($P4mm$ phase) and $[ 111 ]$ ($R3m$ phase) directions. 
At higher pressures the $R3m$ phase was proposed to transform to a cubic $Pm\bar{3}m$ phase. By symmetry, these phases do 
not possess octahedral tiltings: the role of the $Cm$ phase is to be an energetically favourable 
path between the $P4mm$ and $R3m$ phases. 
An alternative path is a commensurate transition involving octahedral tiltings, which are 
frequently observed in oxide perovskites. This was suggested to occur also in PbTiO$_3$ under high pressure  
according to the phase transition sequence $P4mm \rightarrow I4cm \rightarrow I4/mcm \rightarrow I4cm$ 
\cite{Kornev}. The proposed octahedral tilt system is the same as is found in SrTiO$_3$. However, octahedral 
tilting is also found in rhombohedral $R3c$ and $R\bar{3}c$ phases. The $R3c$ space group is a generic rhombohedral 
symmetry in the sense that it allows both octahedral tilting and cation displacements.  Thus, it is important to 
consider both $R3m$ symmetry (only cation displacements are allowed) and its subgroup $R3c$. The latter symmetry 
corresponds to the case where the two oxygen octahedra of the primitive cell are tilted about the pseudocubic 
$[ 111 ]$ direction. The study of phonon instabilities provides invaluable information: 
particularly the phonon frequencies at the high symmetry points of the Brillouin zone and 
the acoustical branch provide very 
useful information related to the possible phase transitions. The first mentioned case 
provides information 
about commensurate phase transition whereas the incommensurate phase is signalled by a dip in the lowest frequency 
acoustical phonon branch. An incommensurate phase often bridges phases which do not possess group-subgroup relationship.

\begin{figure}[p]
\begin{center}
\includegraphics[width=8.6cm,angle=270]{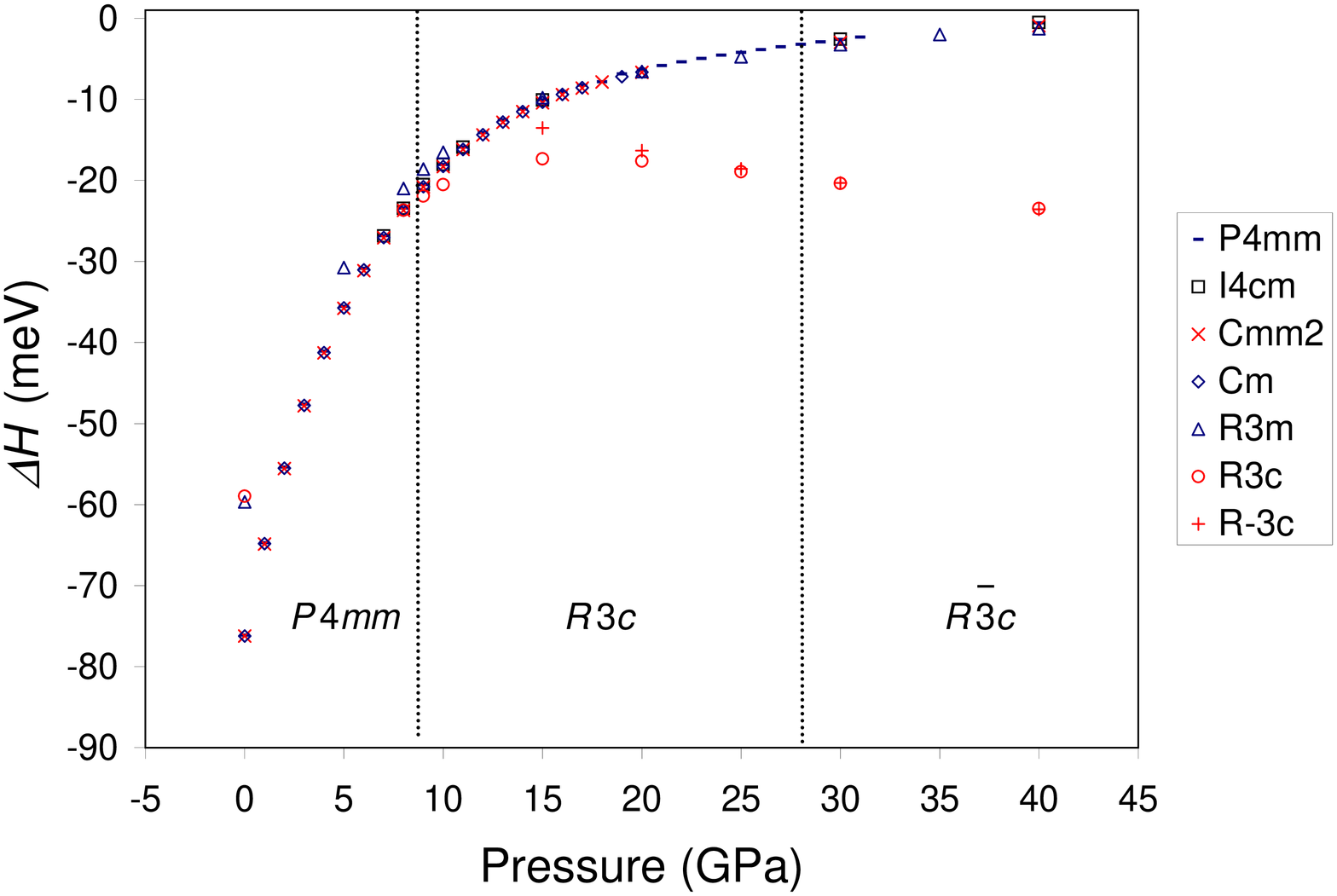}
\caption{\label{Enthalpies} Enthalpies as a function of pressure with respect to the cubic $Pm\bar{3}m$ phase. 
Our first-principles computations reveal that energetically favourable and large crystal compression is allowed 
by tilting the oxygen octahedra. The minimum enthalpy path is $P4mm \rightarrow R3c \rightarrow R\bar{3}c$.}
\end{center}
\end{figure}

\begin{figure}[p]
\begin{center}
\includegraphics[width=8.6cm,angle=270]{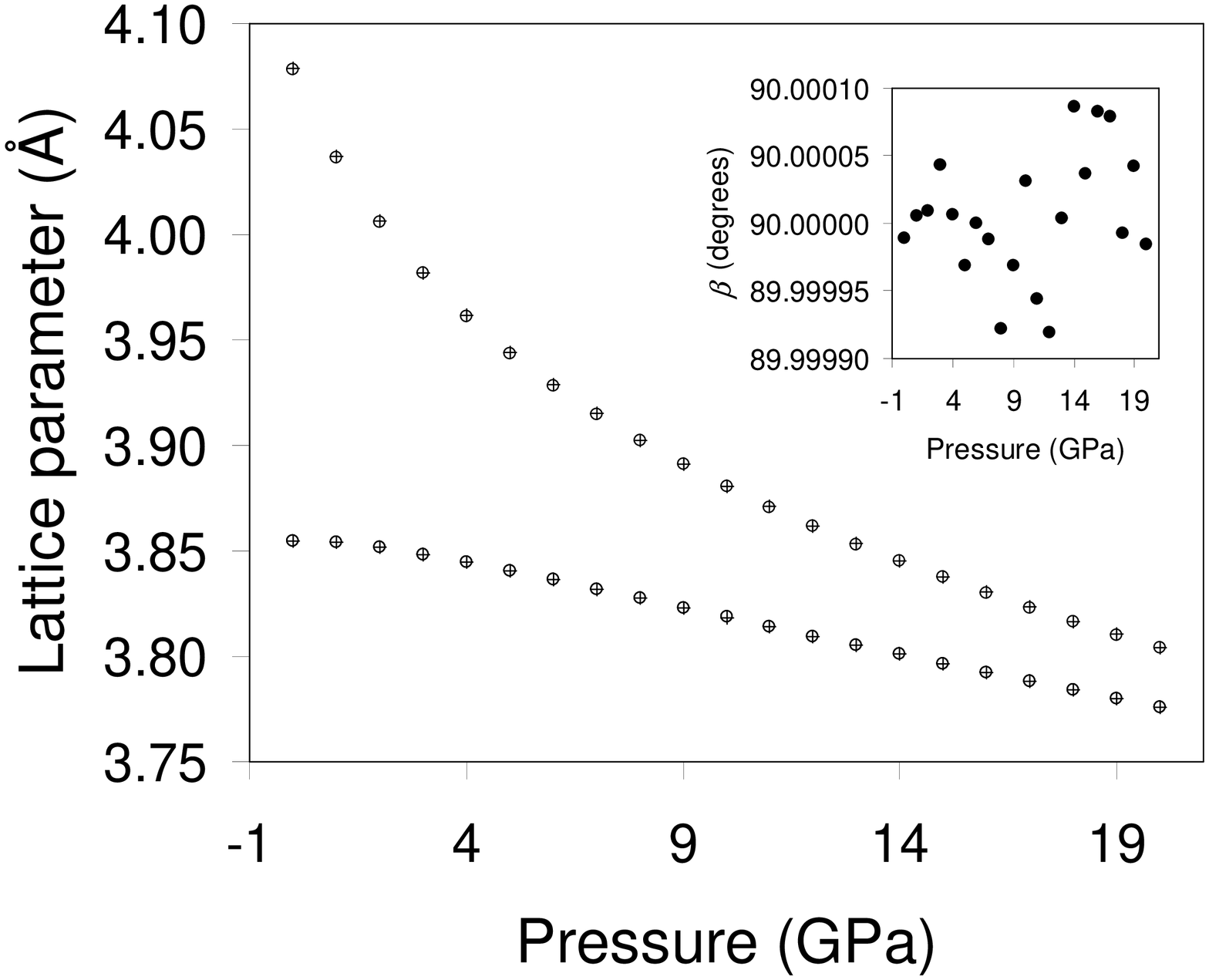}
\caption{\label{LatticeParameters} Lattice parameters for the $P4mm$ (crosses) and $Cm$ 
(open circles) phases. 
For an easier comparison, the monoclinic $a$ and $b$ axes are divided by $\sqrt{2}$. 
Inset shows the behaviour of the monoclinic $\beta$ angle 
as a function of pressure. It is seen that monoclinic distortion is essentially zero 
in the studied pressure range and thus PbTiO$_3$ has $P4mm$ symmetry up to 9 GPa at which 
a transition to the $R3c$ phase takes place.}
\end{center}
\end{figure}

\begin{figure}[p]
\begin{center}
\includegraphics[width=8.0cm,angle=0]{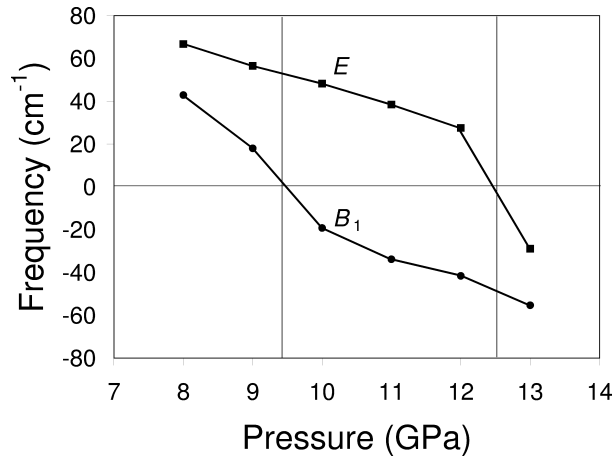}
\caption{\label{B1andEsymmetryModes} Brillouin zone boundary modes at $A$ point 
($\mathbf{q}=(\frac{\pi}{a}\frac{\pi}{a}\frac{\pi}{c}$) 
become unstable at 10 GPa ($B_1$ mode) and at 13 GPa ($E$ symmetry mode). Lines 
are guides for the eyes.}
\end{center}
\end{figure}

\begin{figure}[p]
\begin{center}
\includegraphics[width=6.6cm]{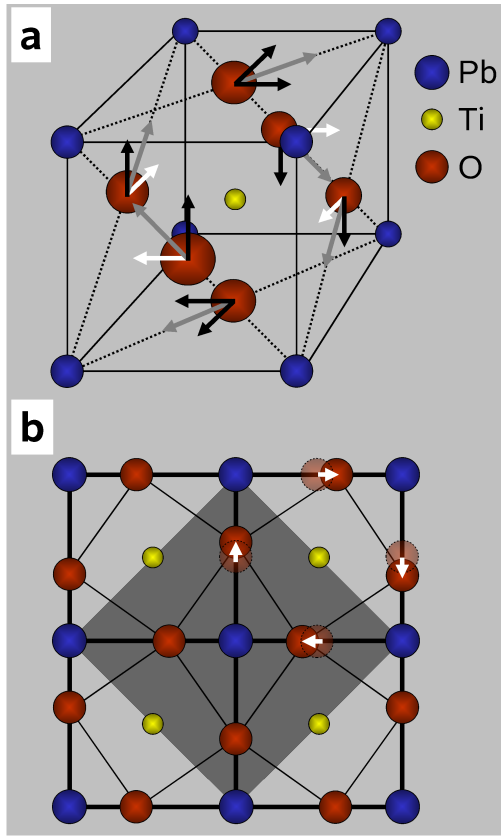}
\caption{\label{PhononInstabilities} Phase transition between $P4mm$ and $R3c$ phases corresponds 
to the condensation of two modes.  
a, Oxygen displacements in two unstable modes corresponding to the 
Brillouin zone boundary $A$ point. White and black arrows indicate the oxygen 
displacements in the $B_1$ and $E$ symmetry modes, respectively. Grey arrows indicate the 
superposition of the $B_1$ and $E$ symmetry modes resulting in rhombohedral symmetry (the 
shown octahedra is rotated clockwise and the next oxygen octahedra (not shown) along the 
$[ 111 ]$ direction is rotated anticlockwise). As a result of the condensation of these 
modes the symmetry discontinuously changes to a rhombohedral $R3c$ symmetry. 
b, Oxygen octahedral tilting in $I4cm$ phase. $B_1$ symmetry mode condensation corresponds to 
the phase transition between $P4mm$ symmetry and its subgroup $I4cm$. The $ab$ basal plane of 
the $I4cm$ unit cell is indicated by dark grey shading.}
\end{center}
\end{figure}

Figure \ref{Enthalpies} shows the enthalpies for $P4mm$, $Cmm2$, $Cm$, $I4cm$, $R3m$ and $R3c$ phases with 
respect to cubic $Pm\bar{3}m$ phase. As far as $P4mm$ and $R3m$ phases are concerned, it bears a close resemblance with  
the enthalpy curves given in ref. \cite{Wu}. The difference between the enthalpies is very small except for the 
rhombohedral symmetries. This is due to the fact that also in the case of the $Cm$ and $Cmm2$ symmetries the structural 
parameters turned out to correspond to the $P4mm$ though orthorhombic and monoclinic distortions were allowed.
Neither orthorhombic nor monoclinic was energetically favourable as Figure \ref{LatticeParameters} shows. Figure \ref{Enthalpies} 
reveals that the minimum enthalpy path corresponds to the phase transition sequence $P4mm \rightarrow R3c \rightarrow R\bar{3}c$.
It is worth mentioning that, within the studied pressure range from 0 to 40 GPa, neither the $Cm$ nor $R3m$ phase was a ground state 
although with increasing pressure the $Cm$ phase became more favourable than $P4mm$ phase and the $R3m$ phase was more favourable 
than the $Cm$ phase at even higher pressures. For the electromechanical properties it is crucial to note that 
the phase transition between the $P4mm$ and $R3c$ phases must be of first order and thus does not involve continuous polarization rotation. 
To put these observations on a firm ground phonon frequencies were computed at the high symmetry points of the Brillouin zone as a function 
of pressure. At zero pressure no phonon instabilities in the $P4mm$ phase were observed, consistently with 
experiments. However, once the pressure is increased the Brillouin 
zone corner ($A$ point, $\mathbf{q}= (\frac{\pi}{a} \frac{\pi}{a} \frac{\pi}{c}$)) $B_1$ mode 
becomes unstable at 10 GPa, Figure \ref{B1andEsymmetryModes}. By further increasing the 
pressure also the $E$ symmetry mode 
becomes unstable at 13 GPa. No other instabilities or anomalies in acoustical branches were observed. 
Though the frequencies of the Brillouin zone centre $A_1$(1TO) and $E$(1TO) modes were decreasing with 
increasing pressure they frequencies remained finite at the whole pressure range.
Figure \ref{PhononInstabilities} a shows the oxygen displacements in $B_1$ and $E$ symmetry
and the structural distortion following the condensation of both modes. This bears a close analogy to 
the $Pm\bar{3}m \rightarrow R\bar{3}c$ phase transition observed in LaAlO$_3$ \cite{Lines}. 
The fact that the $B_1$ mode becomes unstable first suggests that the phase transition may occur in two 
stages. 
The oxygen octahedral rotation corresponding to the condensation of the $B_1$ mode is 
depicted in Figure \ref{PhononInstabilities} b and is similar to the case 
observed in SrTiO$_3$ \cite{Lines}. The second stage is shown by black arrows in Figure 
\ref{PhononInstabilities} a. However, the $I4cm$ symmetry did not correspond to the minimum 
enthalpy, Figure \ref{Enthalpies}. 
This implies that either the stability range of $I4cm$ phase is very narrow or the phase 
transition between $P4mm$ and $R3c$ phase takes place without an intermediate phase. 
In the case of a first-order phase transition the frequencies 
of the unstable modes do not reach zero at the transition point. 
It is remarkable that the information provided by the phonon 
instabilities, most notably by the phonon eigenvectors, is perfectly consistent with the analysis based on enthalpies. 

\begin{figure}[p]
\begin{center}
\includegraphics[width=11.0cm,angle=0]{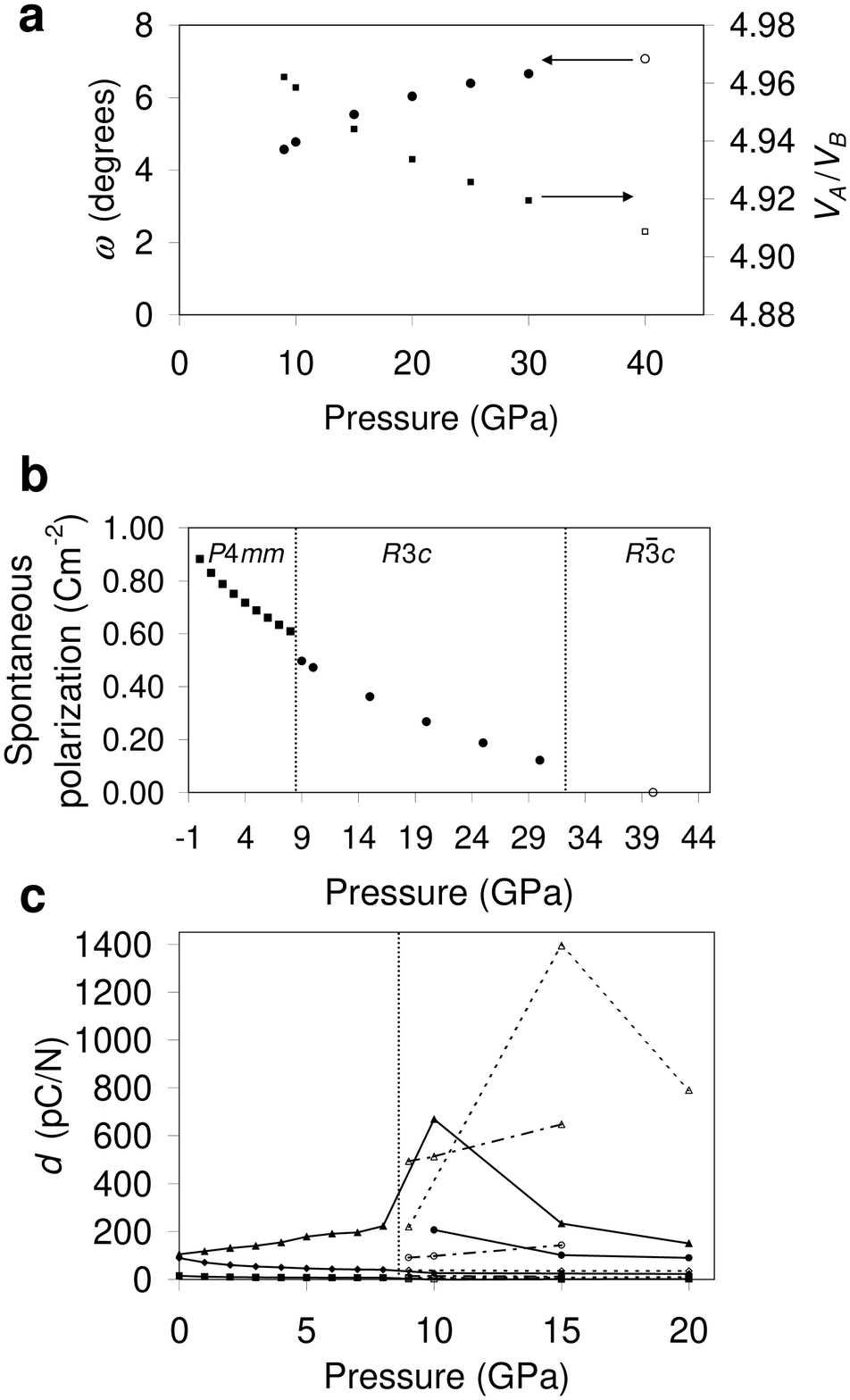}
\caption{\label{PolyhedraVolumes} a, The ratio between cuboctahedra ($V_A$) and oxygen octahedra volume 
($V_B$) as a function of pressure in $R3c$ (filled marks) and $R\bar{3}c$ (open marks) phases. The 
corresponding octahedral tilting angle $\omega$ is also shown.
b, Spontaneous electric polarization across the proposed phase transition sequence. 
c, Absolute values of the piezoelectric constants for the stable phases (continuous line with filled symbols) 
and for the unstable phases ($P4mm$: dashed lines, $R3m$: dashed-dotted line). Squares, diamonds, triangles and 
spheres indicate $d_{31}$, $d_{33}$, $d_{15}$ and $d_{11}$ constants, respectively. The anomalously large 
$d_{15}$ values of the $R3m$ and $P4mm$ phases at 15 GPa pressure are a clear indication of a structural instability. 
Symmetry dictates that for these phases polarization rotation is not possible and thus cannot offer an explanation 
for the anomaly.}
\end{center}
\end{figure}

The $V_A/V_B$ ratio decreases with increasing hydrostatic pressure in rhombohedral $R3c$ and 
$R\bar{3}c$ phases. The central role of the tilt angle of the octahedron about $[ 111 ]$ direction ($\omega$) is neatly 
given by the expression $V_A/V_B \approx 6 \cos^2 \omega-1$ \cite{Thomas}. 
This is related to the increasing octahedral tilting and is clearly an important mechanism for 
minimizing the enthalpy. 
This is seen by comparing Figures \ref{Enthalpies} and \ref{PolyhedraVolumes} \textbf{a} 
and noting that $R3m$ phase does not 
permit octahedral tiltings. Figure \ref{PolyhedraVolumes} \textbf{b} shows that spontaneous 
polarization decreases 
monotonically with increasing pressure. There is a discontinuous drop in spontaneous polarization 
at 9 GPa so that not only is the polarization direction changed abruptly but also the magnitude 
drops significantly. With increasing pressure the cation shifts diminish and a transition 
between the $R3c$ and $R\bar{3}c$ phases occurs at around 30 GPa, see Figure \ref{Enthalpies}. At 30 GPa 
the cation shifts from the centrosymmetric positions are small, which suggests that the transition is 
continuous. These results mean that under high pressure PbTiO$_3$ does not minimize its enthalpy through 
polarization rotation but by tilting the oxygen octahedra. This has very important consequence for the 
piezoelectric properties: though the $d_{15}$ constant has a clear maximum at the phase transition 
there is no dramatic enhancement of piezoelectric constants at the phase transition 
pressures, as Figure \ref{PolyhedraVolumes} \textbf{c} shows. This is in complete contrast to the predictions 
based on the polarization rotation theory \cite{Fu,Wu}: in the pressure range between 9 and 15 GPa 
significantly larger piezoelectric coefficients are seen in the case of the $R3m$ and $P4mm$ phases, which 
do not permit polarization rotation. We conclude that it is the 
phase instability which results in the large piezoelectric constants. In a classical piezoelectric Pb(Zr,Ti)O$_3$ 
system an analogous situation occurs in the vicinity of the morphotropic phase boundary (MPB) where the $Cm$ phase 
coexists with the $R3m$ phase (at room temperature) or $R3c$ phase (at low temperature)\cite{Frantti}. 
Our computations suggest that it is the phase instabilities of the $R3m$ and/or $Cm$ phase(s) which is 
responsible for the large electromechanical coupling coefficients observed at MPB.

Experimentally this zero temperature phase transition sequence $P4mm \rightarrow R3c \rightarrow R\bar{3}c$  
can be verified through 
neutron diffraction experiments. Neutron diffraction allows oxygen positions to be precisely 
located and the superlattice reflections allow the identification of octahedral tiltings. In 
addition, Raman spectroscopy provides useful complementary information: all phases 
have very distinctive spectra. Since the transition between $R3c$ and $R\bar{3}c$ is likely 
to be continuous, Raman scattering data are especially valuable: the Brillouin zone centre normal 
modes (IR and R stand for the infrared and Raman active modes) of $R3c$ and $R\bar{3}c$ 
transform as the representations 
$5A_1 (IR,R) \oplus A_2 \oplus 10E (IR,R) \oplus$ and 
$A_{2g}\oplus E_g (R) \oplus 4 A_{1u} \oplus 5 A_{2u} (IR) \oplus 9E_u (IR)$, 
respectively. Thus, the $R3c$ phase has 15 peaks (excluding accidental degeneracies) whereas 
$R\bar{3}c$ has only one peak.
 
\section*{Acknowledgments}
This project was financially supported by the Academy of Finland (Project Nos 207071 and 207501).
Finnish IT Center for Science is acknowledged for providing computing environment.

\end{document}